# Enhanced Upper Critical Fields in a New Quasi-one-dimensional Superconductor $Nb_2Pd_xSe_5$

Seunghyun Khim[1], Bumsung Lee[1], Ki-Young Choi[1], Byung-Gu Jeon[1], Dong Hyun Jang[1], Deepak Patil[1], Seema Patil[1], Rokyeon Kim[2], Eun Sang Choi[3], Seongsu Lee[4], Jaejun Yu[2] and Kee Hoon Kim[1]

[1]Center for Novel States of Complex Materials Research, Department of Physics and Astronomy, Seoul National University, Seoul 151-747, Republic of Korea

[2]CSCMR, Department of Physics and Astronomy, Seoul National University, Seoul 151-747, Republic of Korea

[3]National High Magnetic Field Laboratory, Florida State University, Tallahassee, Florida 32310, USA

[4]Neutron Science Division, Korea Atomic Energy Research Institute, Daejeon 305-353, Republic of Korea

Email: khkim@phya.snu.ac.kr

**Abstract.** We report a discovery of superconductivity with $T_c$ = 5.5 K in $Nb_2Pd_xSe_5$, in which one-dimensional (1D) Nb-Se chains existing along the *b*-direction hybridize each other to form the conducting *b*-*c** plane. The magnetic susceptibility and specific heat data in both single- and poly-crystals show evidence of bulk superconductivity. The resistivity, Hall coefficient, and magneto-resistance data all indicate the presence of an energy scale $T^*$ = ~ 50 K, which becomes systematically lowered under hydrostatic pressure and competes with the stabilization of superconductivity. Combined with the band calculation results showing the Fermi surfaces with 1D character, we postulate that the energy scale $T^*$ is related to a formation of a density wave or a stabilization of low dimensional electronic structure. The zero temperature upper critical field, $H_{c2}(0)$, of the single crystal is found to be 10.5, 35 and 22 T in the *a'*, *b* and *c**-directions, respectively. While the linearly increasing $H_{c2}(T)$ for $H$ // *c** indicates the multi-band effect, $H_{c2}(0)$ for $H$ // *b* and *c** are found to be much bigger than the BCS Pauli limiting field, 1.84 $T_c$ ~ 9 T. The suppressed Pauli paramagnetic effect points to a possibility of the enhanced spin-orbit scattering related to the low dimensional electronic structure or the presence of heavy elements such as Pd.

# 1. Introduction

The quasi one-dimensional (q1D) electronic structure is a fertile playground to search for novel states in strongly correlated materials. For example, the low dimensional Fermi surfaces are often susceptible to instability of a spin or charge density wave, and also provide a chance to observe unconventional superconductivity including a singlet *d*-wave [1] and a triplet *p*-wave order parameter [2, 3]. Moreover, the paramagnetic one-dimensional metal without Landau quasi-particle excitations, so called Luttinger liquid, is closely associated with the phenomenon of the spin-charge separation [4, 5].

Interesting phenomena can be also observed in the q1D materials upon applying a magnetic field, whose energy scale is comparable to their inter-chain interaction. It has been predicted that when a high magnetic field is applied perpendicular to the most conductive chain, the electron trajectories become smaller than the distance between the chains, resulting in the localized electron motion in two dimensions [6]. This phenomenon, called the field-induced dimensional crossover, was experimentally evidenced by, for instance, a magnetic field induced metal-insulator transition in the inter-chain resistivity of cuprates [7, 8].

For superconducting properties, the low dimensional electronic structure not only brings up a large anisotropy in $H_{c2}$ but also remarkably weakens the orbital pair-breaking effect to make the Pauli limiting effect more important [9]. However, the observed $H_{c2}$ values are often found to be much larger than the BCS Pauli limiting field in many of such low dimensional systems. For example, under the influence of the dominant Pauli limiting effect, a spatially modulated superconducting state, called a Fulde-Ferrel-Larkin-Ovchinnikov (FFLO) state, can be stabilized [10, 11]. The realization of the FFLO state is expected to stabilize the spin singlet state above the Pauli limiting field. Based on the robustness of superconductivity at high magnetic fields, the spin triplet pairing has been argued [12, 13], e.g., in the q1D superconductor $Li_{0.9}Mo_6O_{17}$ and supported by experimental results presenting the observation of a re-entrance of superconductivity [14] and enhanced $H_{c2}$ behaviors [15].

In this article, we report a discovery of superconductivity in a q1D $Nb_2Pd_xSe_5$, which is isomorphic to the recently reported two other sulfide superconductors, $Ta_2Pd_xS_5$ and $Nb_2PdS_5$ [16, 17, 18]. We found a perfect Meissner shielding in the magnetic susceptibility and confirmed bulk superconductivity at $T_c$ = 5.5 K by the specific heat measurements. According to the band structure calculations, the Fermi surface was composed of multiple bands with different dimensional character and carrier-types. In particular, the presence of the sheet-like Fermi surfaces indicated the one-dimensional character and consequently pointed to a possible Fermi surface nesting effect. Related to this, we observed a slope change in the temperature-dependent resistivity and noticeable increases in the temperature-dependent Hall coefficient and magneto-resistance below ~ 50 K. These observations thereby suggest existence of an energy scale compatible with such a low dimensional electronic structure. In addition, the temperatures at which the transport anomalies occurred were gradually reduced with increasing hydrostatic pressure. From the high field transport studies, $H_{c2}(0)$ was determined to be 10.5, 35 and 22 T for the three axes, $a'$, $b$ and $c^*$, respectively. Most importantly, $H_{c2}(0)$ for the conductive chain direction exceeded the BCS Pauli limiting field of ~ 9 T by about four times. Possible scenarios will be discussed for the origin of the enhanced Pauli limiting field.

## 2. Experiment

Polycrystalline samples were grown by the solid state reaction method. Pure powders of Nb (99.8 %), Pd (99.99 %) and Se (99.99 %) were weighted with a ratio of 2 : $x$ : 5 ($x$ = 0.8 - 1.3), ground well in a mortar, and pressurized into the pellet form. All the samples were prepared inside a glove box filled with a pure Ar ($H_2O$ and $O_2$ level < 1 ppm). The pellet was sealed inside a quartz ampule in vacuum and heated at 750 ℃ for 2 - 3 days, and this process was repeated two or three times. Structural properties were investigated by the X-ray and neutron powder diffraction methods. Neutron powder diffraction data were collected over a $2\theta$ range of 0 - 160 ° with a step size of 0.05 ° and $\lambda$ = 1.8367 Å supplied by a Ge(331) single-crystal monochromator on a high resolution powder diffractometer (HRPD) at the HANARO of the Korea Atomic Energy Research Institute (KAERI). The Rietveld analyses were performed on the X-ray and neutron diffraction data. Superconductivity in the polycrystalline $Nb_2Pd_xSe_5$ specimens was found in a broad nominal Pd composition ranging from $x$ = 0.8 to 1.3 ($T_c$ = 3.4 - 5.5 K). It turned out that all the polycrystalline samples with $x$ = 0.8 - 1.3 showed the X-ray diffraction patterns expected for $Nb_2PdSe_5$ without showing any discernible impurity peaks, and the $x$ = 1.2 sample in particular showed the highest $T_c$ ~ 5.5 K (zero resistivity). For the polycrystalline sample, we thus present the results of the nominal $Nb_2Pd_{1.2}Se_5$ samples with the highest $T_c$.

Single crystals were grown from mixed Nb, Pd, and Se powders with a ratio of 4 : 1 : 10, respectively. The mixed powder was then sealed in a quartz ampoule and heated at 850 ℃ for 10 - 15 days. Needle-like single crystals were harvested and its typical size was 50 μm × 50 μm × 500 μm. Electron-dispersive X-ray spectroscopy (EDX) and electron backscattering diffraction study were performed to determine the actual composition and the crystallographic orientations of the single crystals, respectively. The EDX measurement was performed by use of the polycrystalline $Nb_2Pd_{1.2}Se_5$ sample as a standard specimen.

Resistivity and Hall effect measurements were performed by the conventional 4- and 5-probe methods, respectively. Gold wires were attached to the sample with a silver paste (Dupont 4929N™). Current was applied along the $b$-direction in the resistivity measurements of the single crystal. The current and voltage contacts were made to cover around the needle-shaped sample to minimize the mixing with other resistivity components. Magnetic susceptibility was measured with a vibrating sample magnetometer in PPMS$^{TM}$ (Quantum Design, USA). Specific heat data were measured down to 0.1 K with PPMS$^{TM}$ and a dilution refrigerator by using a custom-made program employing the relaxation method with the two relaxation times. The experiments at high magnetic fields were performed up to 35 T, using a resistive magnet at National High Magnetic Field Laboratory, USA. Resistivity under hydrostatic pressure up to 2.2 GPa has been measured in a piston-type hybrid cell with a pressure medium (Daphne 7373). The internal pressure of the cell was calibrated by measuring the $T_c$ of lead.

Density functional theory (DFT) calculations were performed using the Vienna *ab initio* simulation code [19]. It is noted that the structural parameters for the calculated stoichiometric Nb$_2$PdSe$_5$ were obtained by the refinement of the X-ray powder diffraction data measured in the nominal Nb$_2$Pd$_{1.2}$Se$_5$ sample. The projector-augmented-wave method [20] and the exchange correlation functional of the generalized gradient approximation in the Perdew, Burke, and Ernzerhof scheme were employed. The self-consistent total energy and eigenvalues were evaluated with 10 × 40 × 10 Monkhorst-Pack mesh of *k*-points [21], and the cutoff energy for the plane-wave basis set was 400 eV.

## 3. Results
### 3.1 Crystal structure

Figure 1(a) shows the neutron powder diffraction pattern and the Rietveld refinement results for the polycrystalline Nb$_2$Pd$_{1.2}$Se$_5$ at room temperature. Without including any additional magnetic diffraction peaks, we could successfully refine the measured Bragg peaks by a monoclinic (space group *C12/m*) structure with lattice parameters $a$ = 12.788 Å, $b$ = 3.406 Å, $c$ = 15.567 Å, and $\beta$ = 101.63 °. From the independent X-ray powder diffraction measurements and the Rietveld refinement, we could obtain almost similar lattice parameters within one percent.

Figure 1(b) displays several features of the refined crystal structure of Nb$_2$PdSe$_5$: 1) One dimensional stacks of the edge-sharing Nb(1)-Se and face-sharing Nb(2)-Se trigonal prisms form the most conductive chains along the *b*-direction. 2) The Pd(1) sites located at the center of the Se squares also constitute the part of the conductive *b*-chain. 3) The planar connection along the [1 0 2] direction between the two trigonal prisms and the Pd(1)-Se squares results in the (-2 0 1) planes. 4) The Pd(2) site is sandwiched between the (-2 0 1) planes. These structural features turned out to be well-reflected in the facets and the crystallographic orientations of the single crystal; a needle-shaped crystal was grown along the *b*-axis and easily cleaved along the (-2 0 1) plane (figure 1(c)), of which orientation was determined by the electron backscattering diffraction. For convenience, we define hereafter $a'$, $b$ and $c^*$-directions as [-2 0 1], [0 1 0] and [1 0 2], respectively, in a setting of the conventional unit cell.

The actual atomic composition of the single crystal was determined to be $Nb_2Pd_{0.74}Se_{5.11}$ by the EDX measurements. The proximity to ~ 0.75 of the Pd composition implies that nearly one quarter of the Pd sites are vacant. According to the structural features described above, it is likely that the defects are formed more easily in the Pd(2) site than the Pd(1) site because the former has a longer bond length and weaker bonding strength with the neighboring Se. The weaker Pd(2)-Se bonding seems to be also consistent with the cleaved (-2 0 1) plane. Previous reports on $Nb_2Pd_{0.71}Se_5$ and $Ta_2Pd_{0.89}S_5$ have also indicated that there existed point defects in the Pd(2) sites [22, 23]. Therefore, the presence of the Pd(2) defects is likely to be a common feature in this q1D material.

*3.2 Superconducting transition*

Figure 2(a) and (b) present the resistivity, magnetic susceptibility ($4\pi\chi$) data of the polycrystalline sample, all of which provide evidence of the superconducting transition. The resistivity curve starts to drop near 5.9 K and becomes zero at 5.5 K. The magnetic susceptibility measured at $H = 10$ Oe after zero-field-cooling develops the onset of the Meissner effect at 5.5 K and an almost full shielding below 5.0 K, whereas the field-cooled susceptibility exhibits only a small drop, possibly due to a strong flux pinning effect.

Figure 3 shows the electronic specific heat ($C_{el}$) divided by temperature. Fitting the normal state specific heat just above $T_c$ to the form of $C/T = \gamma + \beta T^2$ yielded the Sommerfeld coefficient $\gamma = 15.7$ mJ/mol K$^2$ and a phononic coefficient $\beta = 0.874$ mJ/mol K$^4$. $C_{el}$ was then obtained by subtracting the phonon contribution from the measured specific heat data. It is noticed in figure 3 that the $C_{el}/T$ data display a clear jump, thereby constituting evidence for bulk superconductivity. As the temperature is further lowered, $C_{el}/T$ is almost exponentially suppressed but remains at a finite value of $\gamma_0 = 2.8$ mJ/mol K$^2$ below 0.6 K, which indicates the presence of a non-superconducting metallic region. In this case, the Sommerfeld coefficient of the superconducting volume becomes $\gamma_n = \gamma - \gamma_0 = 12.8$ mJ/mol K$^2$, which corresponds to about 80 % of the total volume. The normalized specific heat jump $\Delta C_{el}/\gamma_n T_c$ is estimated to be 1.63 ± 0.13, of which error bars depend on the choice of a linearly varying region in the $C_{el}/T$ data below $T_c$. The $\Delta C_{el}/\gamma_n T_c = 1.63$ is larger by about 13 % than the weakly-coupled BCS value of 1.43.

Figure 3 also shows the fitting curves of the BCS $\alpha$-model plus $\gamma_0$, in which $\alpha = 2\Delta_0/k_B T_c$ denotes coupling strength, $\Delta_0$ is the size of the superconducting gap, and $k_B$ is the Boltzmann constant [24]. The one-band $\alpha$-model with $\alpha = 4.0$ seems to roughly explain the specific heat jump and exponentially vanishing behavior at low temperatures. Consistent with the enhanced $\Delta C_{el}/\gamma_n T_c$ value, the coupling strength $\alpha = 4.0$ is also larger than the BCS prediction, 3.54. Both results thus suggest that $Nb_2Pd_xSe_5$ is a BCS superconductor located beyond the weak coupling regime.

Although the one-band $\alpha$-model describes the experimental $C_{el}/T$ data reasonably well, several features are still worth being discussed in figure 3. First, we note that the hump feature around 3.5 K cannot be easily explained by the one-band $\alpha$-model. Although the two-band $\alpha$-model was also employed, the hump feature could not be explained without introducing the volume fraction with lower $T_c$. Therefore, the deviation of the data from the fit at ~ 3.5 K might be due to the presence of a superconducting volume fraction with lower $T_c$. To be consistent with this conjecture, specific heat data measured above 2 K in other batches of the polycrystal specimens showed the smaller hump feature near 3.5 K. Second, except the hump feature, we could also fit experimental data better, particularly at low temperatures below 2.5 K upon employing the two-band $\alpha$-model (See, figure 3) [25]. The coupling strengths of each bands were $\alpha_1 = 4.1$ with the relative fraction $x_{rel} = 0.9$ and $\alpha_2 = 2.1$ with the remaining fraction $1 - x_{rel} = 0.1$, respectively. However, the difference between the one-band and the two-band fitting results was not big so that a type of bands with $\alpha \approx 4.0$ seems to dominate the thermodynamic properties in the superconducting state. Third, although the jump and shape of the $C_{el}/T$ curve are roughly consistent with the one-band $\alpha$-model, it should be reminded that another q1D superconductor $Li_{0.9}Mo_6O_{17}$ with a presumably $p$-wave and triplet pairing symmetry has shown a similar consistency with the one-band $\alpha$-model [26]. Thus, the agreement with the $\alpha$-model alone does not necessarily support the $s$-wave pairing symmetry of $Nb_2Pd_xSe_5$.

*3.3 Transport properties in the normal state*

To investigate the normal state transport properties, the magneto-resistance (MR) and the Hall effect were first measured in the polycrystalline samples. At overall temperatures, MR defined as $\Delta\rho_{xx}(H)/\rho_{xx}(0\,T) = (\rho_{xx}(H) - \rho_{xx}(0\,T))/\rho_{xx}(0\,T)$ showed variation nearly in proportion to $H^2$ which is a usual behavior expected in conventional metallic systems (figure 4(a)). On the other hand, its slope showed rather a strong increase below $T^* = $ ~ 50 K. As a result, MR estimated at 9 T started to increase around $T^*$ to reach as high as 6 % at low temperatures while it remained small ($\leq 1.5$ %) above $T^*$ (figure 5(a)). The transverse resistivity ($\rho_{xy}$) (figure 4(b)) showed almost linear variation up to 9 T in a broad temperature window below 300 K, indicating the multiband effect is not clearly observable in the $\rho_{xy}$ data. On the other hand, the resultant Hall coefficient $R_H = \rho_{xy}/H$ (figure 5(a)) clearly showed peculiar temperature-dependence; upon temperature being lowered, $R_H$ decreases continuously to reach a minimum at $T^* = $ ~ 50 K and increases again below $T^*$. Moreover, the Hall mobility ($\mu_H \equiv R_H/\rho_{xx}$) also shows a most steep increase near $T^*$. Consistent with these transport data, $\rho(T)$ data for both single- and poly-crystals showed rather a steep decrease around 50 K so that $d\rho/dT$ curves show a characteristic, broad maximum around $T^*$. In analogy to the heavy fermion materials showing a broad $d\rho/dT$ maximum near a coherent temperature [27, 28], the maximum of $d\rho/dT$ indicates the presence of an energy scale $T^* = \sim 50$ K, below which a new electronic ground state is stabilized. Therefore, all of the three independent transport properties presented in figure 5 coherently support the existence of a crossover from a high to a low temperature metallic phase at $T^*$.

It should be noted in the single crystal that the resistivity along the conductive *b*-direction shows a slight increase below $\sim 20$ K ($d\rho_b/dT < 0$), while that of the polycrystalline sample becomes more or less saturated. It was found that MR of the single crystal becomes negative near the temperature where $d\rho_b/dT < 0$ is observed (See, e.g., the inset of figure 7(c)). Thus, the resistivity upturn may be associated with the weak localization effect, possibly coming from the Pd deficiency. On the other hand, the resistivity upturn along the conductive chain direction is also reminiscent of the case of the q1D $Li_{0.9}Mo_6O_{17}$, in which an insulating ground state was realized below $\sim 30$ K [29, 30]. Therefore, the origin of the insulating behavior in $Nb_2Pd_xSe_5$ should be further investigated to address whether it is one of common features in the q1D materials or it is just an outcome of the atomic defects.

*3.4 Transport properties under hydrostatic pressure*

Resistivity measurements under hydrostatic pressure were performed on another piece of a single crystal showing an even steeper resistivity increase at low temperatures below $\sim 50$ K. As shown in figure 6(a), the resistivity upturn was gradually suppressed upon increasing pressure to induce the metallic ground state down to $T_c$. In each pressure, $T^*$ was determined by an intersection temperature of the two linearly extrapolated lines in the $\rho(T)$ curve below and above the transition and $T_c$ was determined at which 50 % of the normal state resistivity is realized. Figure 6(b) depicts the summarized phase diagrams for $T_c$ and $T^*$ with respect to pressure variation. Upon increasing pressure, $T_c = \sim 4.7$ K at ambient pressure increases with an initial rate of $\sim 1.8$ K/GPa, and then becomes nearly saturated to $\sim 5.8$ K around 2 GPa. On the other hand, $T^* = \sim 50$ K at ambient pressure systematically decreases to 27 K at 2.1 GPa. The evolutions of $T_c$ and $T^*$ suggest that the superconducting ground state competes with the electronic state whose energy scale is characterized by $T^*$.

*3.5 Upper critical fields*

To determine $H_{c2}(T)$ in the single crystal, we measured the temperature-dependent resistivity at fixed magnetic fields below 9 T (not shown here) and the field-dependent resistivity up to 35 T at fixed temperatures (figure 7). To avoid ambiguity due to the surface superconductivity effect and superconducting fluctuation, $H_{c2}$ was chosen as the critical field at which 50 % of the normal resistivity value was achieved in the superconducting transition. Figure 8(a) summarizes the resultant

$H_{c2}(T)$. The $H_{c2}$ slopes near $T_c$ are determined as - 11.5, - 4.78, - 3.70 T/K for $H // b$, $c^*$ and $a'$ respectively. Based on the anisotropic Ginzburg-Landau (GL) relation,

$$\frac{H_{c2}^i}{H_{c2}^j} = \frac{\sqrt{\sigma_i}}{\sqrt{\sigma_j}} = \frac{\sqrt{\rho_j}}{\sqrt{\rho_i}} \qquad (3.1),$$

the anisotropies of $H_{c2}$ near $T_c$ predict the resistivity ratio $\rho_b : \rho_{c^*} : \rho_{a'} = 1 : 5.8 : 9.7$. This result is consistent with the expectation that the charge transfer would be most effective along the Nb-Se conductive chain, i.e., $b$-direction. On the other hand, the transport anisotropy ratio is much smaller than that of the organic superconductors ($1 : 10^3 : 10^5$) [31, 32] but comparable to that of $Li_{0.9}Mo_6O_{17}$ [15, 29, 33].

The extrapolated $H_{c2}$ values at zero temperature, $H_{c2}(0)$, are 10.5, 35 and 22 T for $H // a'$, $b$ and $c^*$, respectively. If $H_{c2}(0)$ is mainly determined by the orbital pair breaking effect originated from kinetic energy of super-current around vortices, a coherence length is given by the GL relation, $H_{c2}^i(0) = \phi_0/2\pi\xi_j(0)\xi_k(0)$, where $\phi_0$ is the flux quantum and $\xi_i(0)$ is the GL coherence length for the $i$-direction. The calculated $\xi(0)$ values representing a characteristic size of a Cooper pairs are 21, 68 and 43 Å for $a'$, $b$ and $c^*$, respectively. Because all of $\xi(0)$ are clearly longer than the lattice constants, this observation indicates that the superconducting order parameter extends over the several conductive chains and planes.

Due to the lack of established theory to describe $H_{c2}(T)$ in the q1D superconductors, we have tried to apply the Werthamer-Helfand-Hohenberg (WHH) formula for an isotropic BCS superconductor in a dirty limit [34]. In figure 8(a), we could fit $H_{c2}(T)$ for $H // b$ with the WHH formula. The Maki parameter $\alpha_M = \sqrt{2}\, H_{c2}^{orb}(0)/H_P^{BCS} = 6.1$ is calculated from the orbital limiting field $H_{c2}^{orb}(0) = -0.69|dH_{c2}/dT|_{Tc}T_c = 39.8$ T and the BCS Pauli limiting field, $H_P^{BCS} = 1.84\, T_c = 9.2$ T. The spin-orbit coupling constant $\lambda_{so}$ was an adjustable parameter to fit the $H_{c2}(T)$ data. The obtained $\lambda_{so} = 55$ is unusually large, and qualitatively indicates that the large spin-orbit scattering is necessary to describe the $H_{c2}(T)$ behavior.

The angular dependent $H_{c2}$, $H_{c2}(\theta)$, was also measured at 0.3 K in the $a'$-$c^*$ and $a'$-$b$ planes as shown in figure 8(b). For a three-dimensional anisotropic superconductor, $H_{c2}(\theta)$ is expressed by the 3D GL model as

$$H_{C2}(\theta) = \frac{H_{c2}(0°)}{\sqrt{\gamma^2 \sin^2\theta + \cos^2\theta}} \quad (3.2)$$

for a given anisotropic ratio $\gamma = H_{c2}(0°)/H_{c2}(90°)$. On the other hand, the 2D Tinkham model is applicable for superconductors when their coherence lengths perpendicular to layers is much shorter than the interlayer spacing. In this case, $H_{c2}(\theta)$ is expressed as [35]

$$\left|\frac{H_{c2}(\theta)\sin\theta}{H_{c2}(0°)}\right| + \left(\frac{H_{c2}(\theta)\cos\theta}{H_{c2}(90°)}\right)^2 = 1. \quad (3.3)$$

where $\theta$ denotes a tilted angle away from a 2D plane. We found that the observed $H_{c2}(\theta)$ curves are well explained by the 3D GL model with the anisotropy ratio 3.35 and 2.15 for the $a'$-$b$ and $a'$-$c^*$ plane, respectively. However, the cusp-like behavior predicted by the 2D Tinkham model is not consistent with the measured data. Therefore, both the estimated coherence lengths and the $H_{c2}(\theta)$ features suggest that the superconducting order parameter is coherent across the several conductive chains. As a consequence, the field-induced dimensional crossover that allows the Cooper pairs to sustain the orbital pair breaking effect is not likely to occur in this material.

## 4. Discussions

### 4.1 Multiband superconductivity

The band calculations were performed for the stoichiometric compound although the actual single- and poly-crystals might have excessive or deficient Pd. Because the Pd deficiency or excess can be easily generated in the weakly bound Pd(2) sites sandwiched between the main $Nb_4PdSe_{10}$ slabs, the population change in the Pd(2) site is not expected to modify the topology of the band structure significantly but only shift the Fermi energy level. Therefore, it is likely that our band calculations for the stoichiometric compound could capture an essential feature of the electronic structure in $Nb_2Pd_xSe_5$.

Figure 9 depicts the calculated Fermi surfaces in a conventional reciprocal unit cell. The contribution of the $d$-orbitals of Nb and Pd yields the multi-sheet Fermi surfaces consisting of a set of electron-like flat sheets, closed pockets, and hole-like corrugated cylinders. Such characteristic Fermi sheets and cylinders are related to the charge transfers along the 1D Nb-Se chain and the conducting plane on the $Nb_4PdSe_{10}$ slab, respectively. The multiple Fermi surfaces also suggest the possible stabilization of multiple superconducting gaps in $Nb_2Pd_xSe_5$. The linearly increasing $H_{c2}$ for $H // c^*$ down to the lowest temperature can be regarded as an experimental hallmark of the multiband superconductivity, as similarly observed in $MgB_2$ and iron-based superconductors [36, 37].

*4.2 The new energy scale T\* and competition with the superconductivity*
The resistivity upturn behavior observed in the single crystal may indicate the weak localization effect — a quantum interference of conducting electrons due to defects or disorder [38]. However, regardless of the resistivity upturn, both single- and poly-crystal data consistently show the resistivity slope change near $T^*$. Moreover, the weak localization effect, possibly coming from the Pd defects, cannot easily account for the sharp superconducting transition observed in both single- and poly-crystals and the systematic changes of the Hall coefficient/mobility and MR below $T^*$ observed in the polycrystal. Furthermore, the observation of the systematic suppression of $T^*$ and enhancement of $T_c$ with pressure suggests that the low temperature metallic state is clearly competing with superconductivity. The hydrostatic pressure generally increases an electron hopping and leads to an effective 3D transport. Therefore, it is expected that the electronic state stabilized below $T^*$ should have a low dimensional Fermi surface topology, such as a sheet or cylinder-like shape.

One appealing scenario for explaining the observed anomalies is to assume a q1D Fermi surface (sheet-like) which is already formed far above $T^*$ and whose effective Fermi energy is quite small. The Femi surface may then allow a coherent electron transport below $T^*$ while the transport at higher temperatures can be incoherent. In this scenario, the q1D Fermi surface is not likely to favor superconductivity because the superconducting state is competing with the energy scale of $T^*$. Superconductivity in $Nb_2Pd_xSe_5$ is then expected to favor the 2D or 3D like Fermi surface topology.

An alternative scenario for the energy scale $T^*$ would be a gradual stabilization of a weak charge density wave (CDW) order. Note that the crystal structure of $Nb_2Pd_xSe_5$ possesses the same 1D Nb-Se chain as in $NbSe_3$, a renowned CDW material. The band structure calculations of $Nb_2Pd_xSe_5$ consistently predict the parallel sheets of the Fermi surfaces, of which nesting feature may naturally allow the density wave formation. The systematic suppression of $T^*$ and enhancement of $T_c$ with pressure may then indicate that the density wave order gets weaker while the superconducting order gets stronger. The competition between the density wave order and superconductivity has been often observed in low dimensional superconductors. The role of the applied pressure is then expected to change the Fermi surface shape into a more 3D like one to destabilize a nesting feature required to form a density wave order.

On the other hand, it should be noticed that the observed anomalies near $T^*$ does not clearly show a localization feature observed in conventional CDW materials, in which the resistivity upturn or a clear sign changing behavior of Hall coefficients are observed [39]. Moreover, the thermodynamic evidence for the density wave transition such as a specific heat jump near $T^*$ has not been observed so far. In this context, we postulate that the density wave order, even if it exists, might have rather an incommensurate and short-ranged character as indicated by the transport properties in incommensurate CDW systems [40, 41].

*4.3 Suppressed spin paramagnetic effect: unusually high $H_{c2}$*

When a Fermi surface of a superconductor has a strong electronic anisotropy, the orbital pair-breaking effect becomes quite weak in a magnetic field applied along highly conductive directions, giving rise to a large $H_{c2}$. In such a case, the usual Pauli limiting effect, originated from the competition between the Zeeman energy and the superconducting condensation energy in a Cooper pair, should be considered. This behavior is expected from the Chandrasekhar-Clogston relation [42],

$$\frac{1}{2}(\chi_N - \chi_S)H_P^2 = \frac{1}{2}N(E_F)\Delta_0^2 \qquad (4.1),$$

where $\chi_N$ ($\chi_S$) is a spin susceptibility of a normal (superconducting) state, $N(E_F)$ is a density of state at the Fermi level and $\Delta_0$ is a superconducting gap at 0 K. In a weakly coupled BCS limit, this relation provides the Pauli limiting field as $H_P^{BCS} = 1.84\,T_c$. Remarkably, all $H_{c2}(0)$ are larger than $H_P^{BCS}$ (= 9.2 T) in $Nb_2Pd_xSe_5$. In particular, $H_{c2}(0)$ for $H // b$ reaches to 35 T, which is nearly 4 times larger than $H_P^{BCS}$. Therefore, it remains to be understood how superconductivity is robust against the spin paramagnetic effect. It is also worthwhile to remind that the unusually high $H_{c2}(0)$ beyond $H_P^{BCS}$ have been found in many other q1D organic and inorganic materials so that the phenomenon seems to be quite generic.

Equation (4.1) suggests that the superconducting gap enhancement due to the strong coupling effect may give rise to a large $H_{c2}$ beyond $H_P^{BCS}$ [43]. Although the observed specific heat jump implies a strong-coupling superconductor ($\alpha \approx 4.0$), the enhancement of 13 % over the BCS limit ($\alpha \approx 3.54$) is insufficient to explain the 4-fold enhancement of $H_{c2}$. Moreover, an unusual enhancement of $H_{c2}(0)$, often observed in the FFLO state due to a spatial modulation of a superconducting order parameter, is not likely to occur here because the hysteresis expected in the first order transition was not observed and the shapes of our $H_{c2}(T)$ curves is almost saturated along both $H // b$ and $H // c^*$ directions.

One explanation for the enhanced Pauli limiting field is the enhancement of the spin-orbit scattering. As implied by (4.1), the finite $\chi_s$ generated by the spin-flip scattering should reduce the Zeeman energy term and result in a larger $H_P$ than $H_P^{BCS}$. This mechanism has been widely suggested to explain the observation of the large $H_{c2}(0)$ above $H_P^{BCS}$ in the superconductors containing a large Z number atom [43, 44]. The spin-orbit scattering becomes a significant portion of the transport scattering for the large Z number as known from the Abrikosov relation $\tau_{tr}/\tau_{so} = (Z/137)^4$ where $1/\tau_{tr}$ ($1/\tau_{so}$) is transport (spin-orbit) scattering rate, which reaches as high as 0.013 for $Z_{Pd} = 46$ [45].

By using the Drude model with the measured Sommerfeld coefficient and the single crystal resistivity value, we estimated that the mean free path ($l_{tr}$) is ~ 5 Å and $1/\tau_{tr}$ is ~ $10^{14}$ sec$^{-1}$ [46]. This estimation implies that this superconductor is in a dirty limit ($l_{tr} < \xi_b = 68$ Å). Furthermore, based on the spin-orbit scattering constant $\lambda_{so}=2\hbar/3\pi k_B T_c \tau_{so}$ obtained from the fit in figure 8(a), $1/\tau_{so}$ becomes ~$10^{12}$ sec$^{-1}$, which is about 100 times smaller than $1/\tau_{tr}$. Therefore, although it is valid for one-band isotropic superconductors, the fit in figure 8(a) seems to satisfy the WHH assumption ($1/\tau_{so} < 1/\tau_{tr}$) and the ratio of $\tau_{tr}/\tau_{so}$ seems to roughly meet the Abrikosov relation. Therefore, the large $H_{c2}$ might arise from the presence of large spin-orbit scattering in a dirty-limit superconductor, particularly related with the presence of ions with a large Z number such as Pd.

Although the presence of Pd or its deficiency might be a main origin to cause the enhanced $\lambda_{so}$ in this system, it seems still worthwhile to consider a mechanism based on the common feature of such q1D electronic structure because the greatly enhanced $\lambda_{so}$ within the WHH scheme seems to be generic in many q1D superconductors [15, 47]. In this context, an alternative route to increase the spin-orbit scattering is the breaking of inversion symmetry, which might be associated with an inherent nature of the low dimensional electronic structure [48]. In thin film superconductors, the inversion symmetry breaking at the interface is known to result in the Rashba-type spin-orbit coupling and gives rise to the enhanced $H_{c2}$ beyond $H_P^{BCS}$ as well as the anomalously enhanced MR [49, 50]. In the q1D materials like NbSe$_3$, an incommensurate CDW introduces the breaking of inversion symmetry [51, 52]. In analogy with those q1D systems, the density wave order, possibly stabilized below $T^*$ in Nb$_2$Pd$_x$Se$_5$ might play a role in increasing the spin-orbit coupling. If this scenario is valid, a local inversion symmetry breaking induced by a CDW-like order should not only result in an enhanced $H_{c2}$ but also point to a stabilization of a peculiar superconducting ground state with an admixture of the spin singlet and spin triplet component [12, 53]. Such an exotic superconducting state has been claimed to exist in other q1D superconductors such as (TMTSF)$_2$PF$_6$ and Li$_{0.9}$Mo$_6$O$_{17}$ [2, 15]. It should be also noted that those two superconductors share a common feature of the large $H_{c2}(0)$ beyond the conventional Pauli limiting field as observed in this study. Therefore, understanding the nature of a possible density wave order and searching for evidence for the spin triplet pairing state seem to be all important ingredients in elucidating the true origin of the greatly enhanced $H_{c2}(0)$ with large spin-orbit scattering in Nb$_2$Pd$_x$Se$_5$.

## 5. Conclusion

In conclusion, we discovered superconductivity in $Nb_2Pd_xSe_5$ with $T_c$ as high as 5.9 K, in which q1D Nb-Se bonding forms the conductive chain along the $b$-direction. The bulk superconductivity was confirmed by the magnetic susceptibility and specific heat data in both single- and poly-crystals. The commonly observed anomalies at $T^* = \sim 50$ K in various transport quantities and its negative pressure dependence imply the existence of a new energy scale that is compatible with the low dimensional electronic structure. $H_{c2}(0)$ of the single crystal in the magnetic field applied along the most conductive $b$-direction is found to be 35 T, greatly exceeding the BCS Pauli limiting field $H_P^{BCS} = 1.84\ T_c \sim 9$ T . The enhanced spin-orbit scattering inherent to the heavy ion elements or to the low dimensional system have been discussed as possible origins to explain the large $H_{c2}(0)$. A possibility to observe the superconducting ground state with mixed spin triplet and single states has been also discussed if the inversion symmetry is locally broken in the low dimensional superconductors.

**Note added**

After completing the work and while writing the manuscript, we became aware that similar superconducting and upper critical field properties were reported in the isostructural compounds $Nb_2Pd_xS_5$ with $T_c < 6.6$ K [17, 54]. Zhang *et al* [18] also reported the observation of superconductivity in $Nb_2Pd_xSe_5$ ($0.67 \leq x \leq 0.7$) with $T_c$ ranging from less than 0.3 K to ~ 1.0 K. It should be noted that $T_c$ values of those compounds are much lower than the maximum $T_c \sim 5.5$ K in the present work. Furthermore, after submission of the present manuscript, we also become aware that another preprint on $TaPd_xS_5$ has appeared [16], reporting $T_c \sim 6$ K and a similar observation of greatly enhanced upper critical fields.

**Acknowledgement**

The authors are grateful to Seon Joong Kim for his critical reading of manuscript and Ju-Young Park and Suk Ho Lee for the EDX measurements. This work was financially supported by the National Creative Research Initiative (2010-0018300) through the NRF of Korea funded by the Ministry of Education, Science and Technology, and the Basic Science Research Program (2012-008233) funded by the Korean Federation of Science and Technology Societies. Work at NHMFL was performed under the auspices of the NSF, the State of Florida, and the U. S. DOE.

**References**

[1] Shinagawa J, Kurosaki Y, Zhang F, Parker C, Brown S E, Jérome D, Christensen J B and Bechgaard K 2007 *Physical Review Letters* **98** 147002

[2]  Lee I J, Naughton M J, Danner G M and Chaikin P M 1997 *Physical Review Letters* **78** 3555-3558
[3]  L. P. Gor'kov D J 1985 *J Phys Lett (Paris)* **46** L643
[4]  Haldane F D M 1981 *Journal of Physics C: Solid State Physics* **14** 2585-2609
[5]  I. E. Dzyaloshinskii A I L 1974 *Sov Phys JETP* **38** 202
[6]  Lebed A G 1986 *Sov Phys JETP* **44** 114
[7]  Hussey N E, Kibune M, Nakagawa H, Miura N, Iye Y, Takagi H, Adachi S and Tanabe K 1998 *Physical Review Letters* **80**, 2909-2912
[8]  Nussey N E, McBrien M N, Balicas L, Brooks J S, Horii S and Ikuta H 2002 *Physical Review Letters* **89**, 086601
[9]  Dupuis N and Montambaux G 1994 *Physical Review B* **49** 8993-9008
[10] Fulde P and Ferrell R A 1964 *Physical Review* **135** A550-A563
[11] A. I. Larkin Y N O 1965 *Sov Phys JETP* **20** 762
[12] Gor'kov L P and Rashba E I 2001 *Physical Review Letters* **87** 037004
[13] Abrikosov A A 1983 *Journal of Low Temperature Physics* **53** 359-374
[14] Xu X, Bangura A F, Analytis J G, Fletcher J D, French M M J, Shannon N, He J, Zhang S, Mandrus D, Jin R and Hussey N E 2009 *Physical Review Letters* **102** 206602
[15] Mercure J F, Bangura A F, Xu X, Wakeham N, Carrington A, Walmsley P, Greenblatt M and Hussey N E 2012 *Physical Review Letters* **108** 187003
[16] Lu Y F, Takayama T, Bangura A F, Katsura Y, Hashizume D and Takagi H 2013 arXiv:1308.3766
[17] Zhang Q, Li G, Rhodes D, Kiswandhi A, Besara T, Zeng B, Sun J, Siegrist T, Johannes M D and Balicas L 2013 *Scientific Reports* **3** 1446
[18] Zhang Q, Rhodes D, Zeng B, Besara T, Siegrist T, Johannes M D and Balicas L 2013 *Physical Review B* **88** 024508
[19] Kresse G and Furthmüller J 1996 *Physical Review B* **54** 11169-11186
[20] Blöchl P E 1994 *Physical Review B* **50** 17953-17979
[21] Monkhorst H J and Pack J D 1976 *Physical Review B* **13** 5188-5192
[22] Keszler D A, Ibers J A, Shang M and Lu J 1985 *Journal of Solid State Chemistry* **57** 68-81
[23] Squattrito P J, Sunshine S A and Ibers J A 1986 *Journal of Solid State Chemistry* **64** 261-269
[24] Padamsee H, Neighbor J E and Shiffman C A 1973 *Journal of Low Temperature Physics* **12** 387-411
[25] Bouquet F, Wang Y, Fisher R A, Hinks D G, Jorgensen J D, Junod A and Phillips N E 2001 *Europhysics Letters* **56** 856-862
[26] Lebed A G and Sepper O 2012 arXiv:1211.1961
[27] Kim K H, Harrison N, Jaime M, Boebinger G S and Mydosh J A 2003 *Physical Review Letters* **91** 256401
[28] Gegenwart P, Custers J, Geibel C, Neumaier K, Tayama T, Tenya K, Trovarelli O and Steglich F 2002 *Physical Review Letters* **89** 056402
[29] da Luz M S, dos Santos C A M, Moreno J, White B D and Neumeier J J 2007 *Physical Review B* **76** 233105
[30] Chen H, Ying J J, Xie Y L, Wu G, Wu T and Chen X H 2010 *Europhysics Letters* **89** 67010
[31] Ishiguro T, Yamaji K and Saito G 1998 *Organic Superconductors* (Berlin: Springer)
[32] Lebed A G 2008 *The Physics of Organic Superconductors and Conductors* (Berlin: Springer)
[33] dos Santos C A M, White B D, Yu Y K, Neumeier J J and Souza J A 2007 *Physical Review Letters* **98** 266405
[34] Werthamer N R, Helfand E and Hohenberg P C 1966 *Physical Review* **147** 295
[35] Tinkham M 1963 *Physical Review* **129** 2413-2422
[36] Gurevich A 2003 *Physical Review B* **67** 184515
[37] Khim S, Lee B, Kim J W, Choi E S, Stewart G R and Kim K H 2011 *Physical Review B* **84** 104502
[38] Lee P A and Ramakrishnan T V 1985 *Review of Modern Physics* **57** 287
[39] Ong N P and Brill J W 1978 *Physical Review B* **18** 5265
[40] Lue C S, Sivakumar K M and Kuo Y K 2007 *Journal of Physics: Condensed Matter* **19** 406230

[41] Kuo Y K, Chen Y Y, Wang L M and Yang H D 2004 *Physical Review B* **69** 235114
[42] Clogston A M 1962 *Physical Review Letters* **9** 266-267
[43] Neuringer L J and Shapira Y 1966 *Physical Review Letters* **17** 81-84
[44] Bauer E, Hilscher G, Michor H, Paul C, Scheidt E W, Gribanov A, Seropegin Y, Noël H, Sigrist M and Rogl P 2004 *Physical Review Letters* **92** 027003
[45] Abrikosov A A and Gor'kov L P 1962 *JETP Letters* **15** 752
[46] Ashcroft N W and Mermin N D 1976 Solid State Physics (New York: Holt Rinehart and Winston) pp 2-49. We used the relation, $1/\tau_{tr} = (\rho[\mu\Omega\ cm]/0.22)(r_s/a_0)^{-3} \times 10^{14}$ [sec$^{-1}$] to estimate the transport scattering rate with the measured $\rho = 3 \times 10^{-3}$ [$\Omega$ cm]. The $r_s/a_0$ term was dropped out by using $\gamma_n = 0.7098Z(r_s/a_0)^2 \times 10^{-4}$ [J/mol K$^2$] and the measured $\gamma_n = 12.8 \times 10^{-3}$ [J/mol K$^2$] where $Z$ is valence ($Z$ is assumed to be a unity), $r_s$ is the electron density and $a_0 = \hbar/me^2$ is the Bohr radious. The mean free path, $l_{tr}$, was obtained from the relation, $l_{tr} = \langle v_F \rangle \times \tau_{tr}$ where $\langle v_F \rangle = 4.20/(r_s/a_0) \times 10^8$ [cm/sec].
[47] Lee I J, Chaikin P M and Naughton M J 2000 *Physical Review B* **62** R14669-R14672
[48] Yip S K 2002 *Physical Review B* **65** 144508
[49] Wu X S, Adams P W, Yang Y and McCarley R L 2006 *Physical Review Letters* **96** 127002
[50] Caviglia A D, Gabay M, Gariglio S, Reyren N, Cancellieri C and Triscone J M 2010 *Physical Review Letters* **104** 126803
[51] Monçeau P, Ong N P, Portis A M, Meerschaut A and Rouxel J 1976 *Physical Review Letters* **37** 602-606
[52] Ong N P and Monceau P 1977 *Physical Review B* **16** 3443-3455
[53] Shimahara H 2000 *Physical Review B* **62** 3524-3527
[54] Niu C Q, Yang J H, Li Y K, Chen B, Zhou N, Chen J, Jiang L L, Chen B, Yang X X, Cao C, Dai J and Xu X 2013 *Physical Review B* **88** 104507

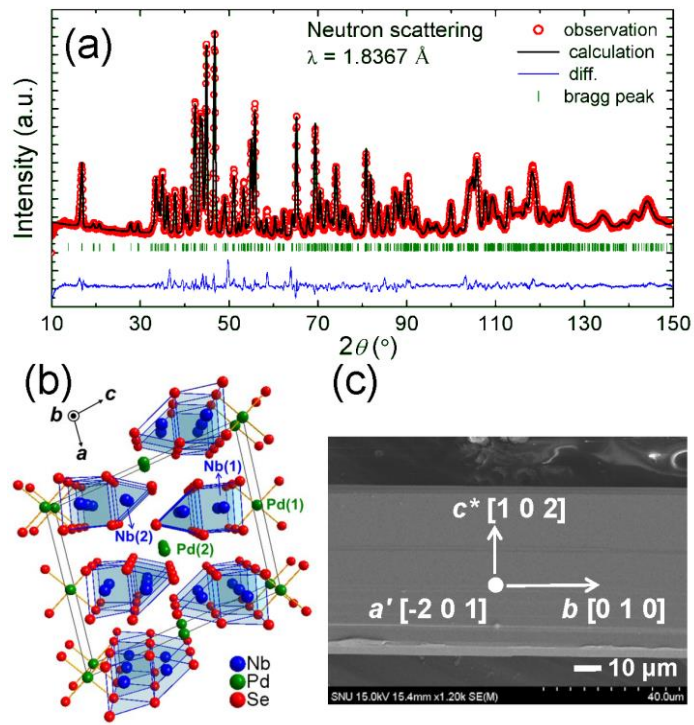

Figure 1. (Color online) (a) The Rietveld refinement result of the powder neutron diffraction data in the polycrystalline $Nb_2Pd_{1.2}Se_5$. (b) The crystal structures of $Nb_2PdSe_5$. (c) The scanning electron microscope image of the $Nb_2Pd_{0.74}Se_5$ single crystal and the crystallographic directions identified by the electron back-scattering diffraction.

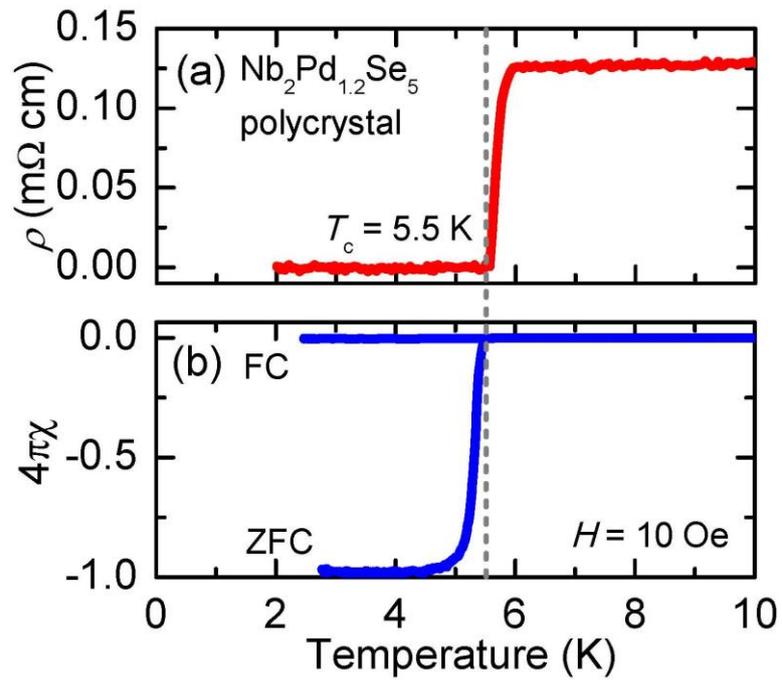

Figure 2. (Color online) The temperature dependence of (a) the resistivity and (b) DC magnetic susceptibility ($4\pi\chi$) in the Nb$_2$Pd$_{1.2}$Se$_5$ polycrystal.

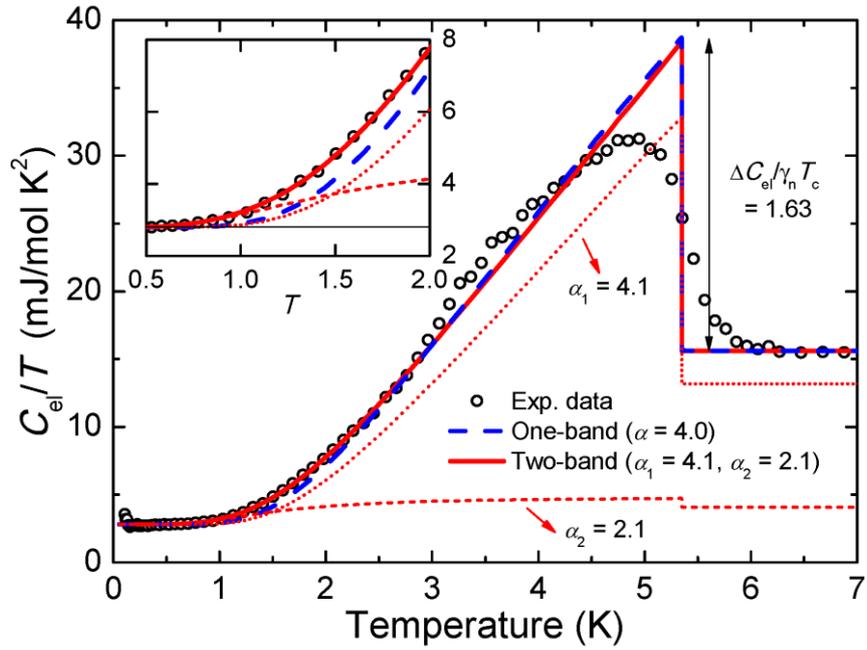

Figure 3. (Color online) The $C_{el}/T$ data in the Nb$_2$Pd$_{1.2}$Se$_5$ polycrystal. The dashed line (blue) and solid line (red) denote the BCS α-model curves for the one-band ($\alpha = 4.0$) and the two-band ($\alpha_1 = 4.1$ ($x_{rel} = 0.9$) and $\alpha_2 = 2.1$ (1-$x_{rel}$ = 0.1)), respectively. Inset: The extended view of $C_{el}/T$ below 2 K. The horizontal line at $C_{el}/T = \gamma_0 = 2.8$ mJ/mol K$^2$ corresponds to the estimated non-superconducting fraction.

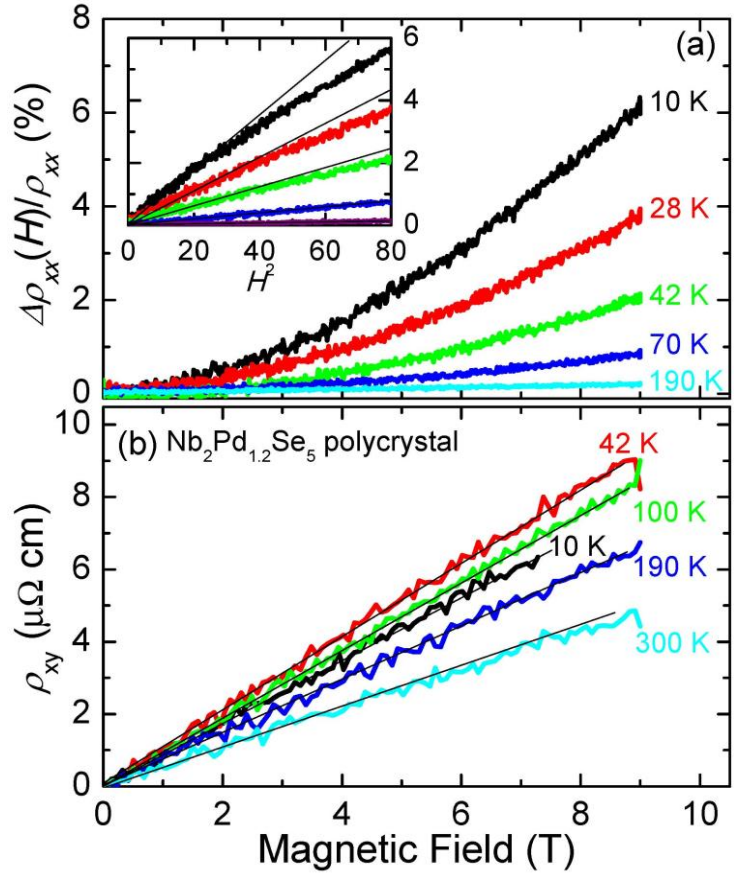

Figure 4. (Color online) (a) The magnetic field dependence of normalized resistivity $\Delta\rho_{xx}(H)/\rho_{xx}(0\,T)$ in the $Nb_2Pd_{1.2}Se_5$ polycrystal. Inset: $\Delta\rho_{xx}(H)/\rho_{xx}(0\,T)$ versus $H^2$. (b) The magnetic field dependence of the Hall resistivity $\rho_{xy}$ in various fixed temperatures. The thin solid lines are linear guides to the eye.

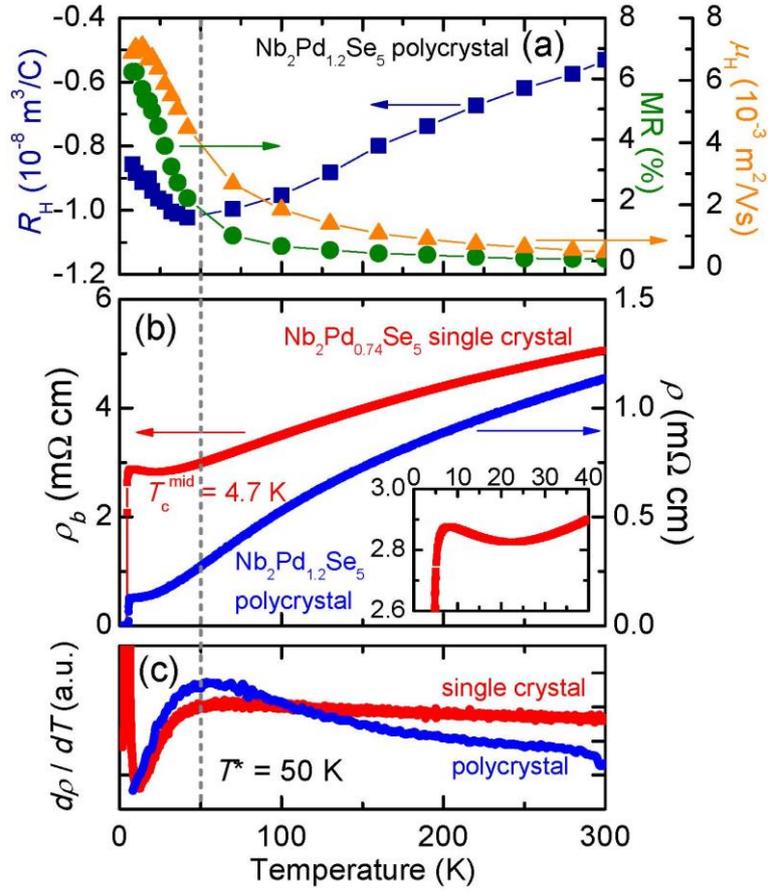

Figure 5. (Color online) (a) The Hall coefficient (dark blue square), Hall mobility (orange triangle) and 9 T magneto-resistance (green circle) measured in the polycrystal. (b) The temperature-dependence of the resistivity of the single crystal (red) and the polycrystal (blue). Inset: The expanded view of the single crystal resistivity in the normal state above $T_c$. (c) The resistivity derivatives with respect to temperature in the single crystal (red) and the polycrystal (blue).

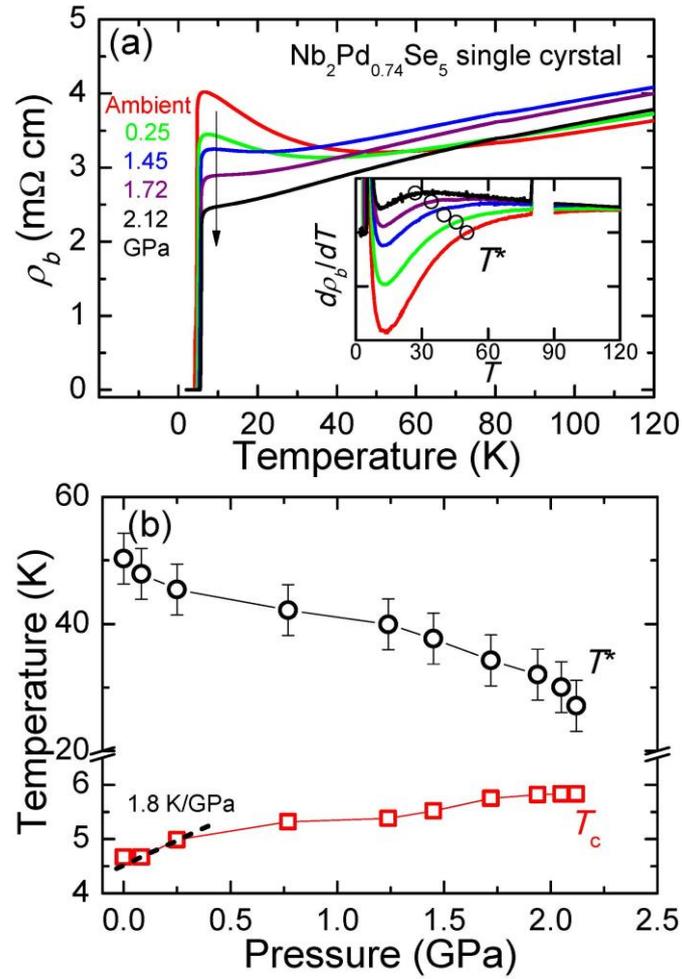

Figure 6. (Color online) The temperature dependence of the single crystal resistivity under pressures of ambient pressure, 0.25, 1.45, 1.72, and 2.12 GPa. Inset: $d\rho/dT$ curves. The open circles denote the anomaly temperatures $T^*$. (b) The resultant phase diagrams of $T^*$ (open black circle) and $T_c$ (open red square) with pressure.

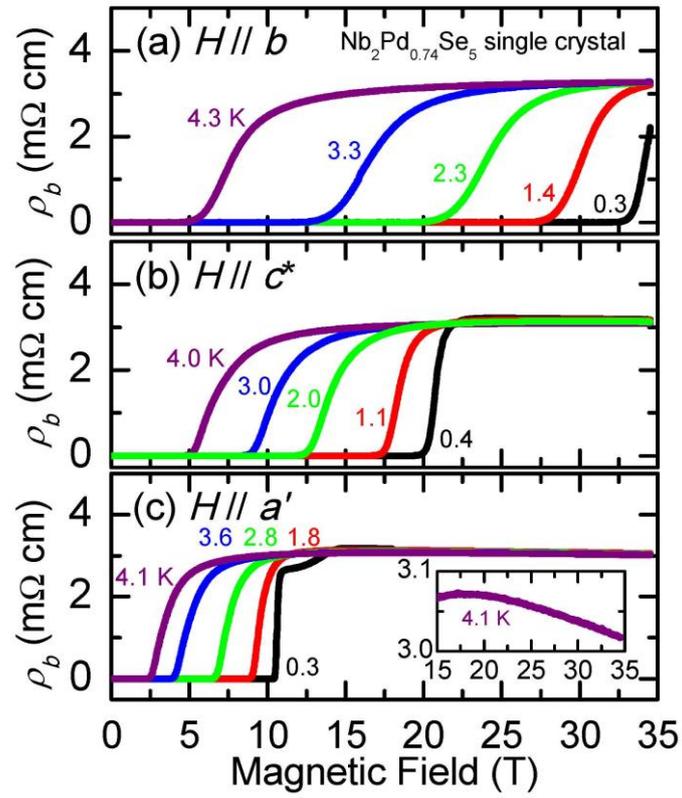

Figure 7. (Color online) The magnetic field dependence of the single crystal resistivity at various fixed temperatures for (a) $H // b$, (b) $H // c^*$ and (c) $H // a'$. The inset shows an expanded view of the magneto-resistance at 4.1 K.

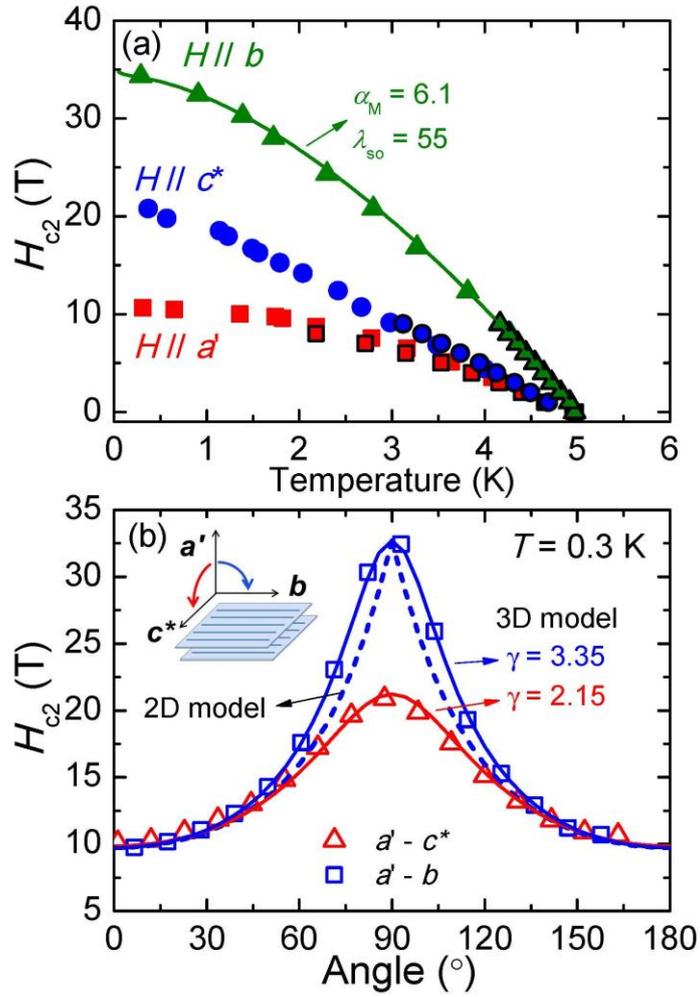

Figure 8. (Color online) (a) The $H_{c2}$ phase diagrams for $H // a'$ (red square), $b$ (green triangle) and $c^*$ (blue circle) -direction in the single crystal. The solid line for $H // b$ shows the WHH curves with the Maki parameter $\alpha_M = 6.1$ and spin-orbit scattering constant $\lambda_{so} = 55$. (b) The angular dependence of $H_{c2}$ measured at 0.3 K in the $a'$-$b$ (open blue square) and $a'$-$c^*$ (open red triangle) plane. The solid lines and the dotted line are depicted by the anisotropic 3D Ginzburg-Landau model and the 2D Tinkham model [35], respectively.

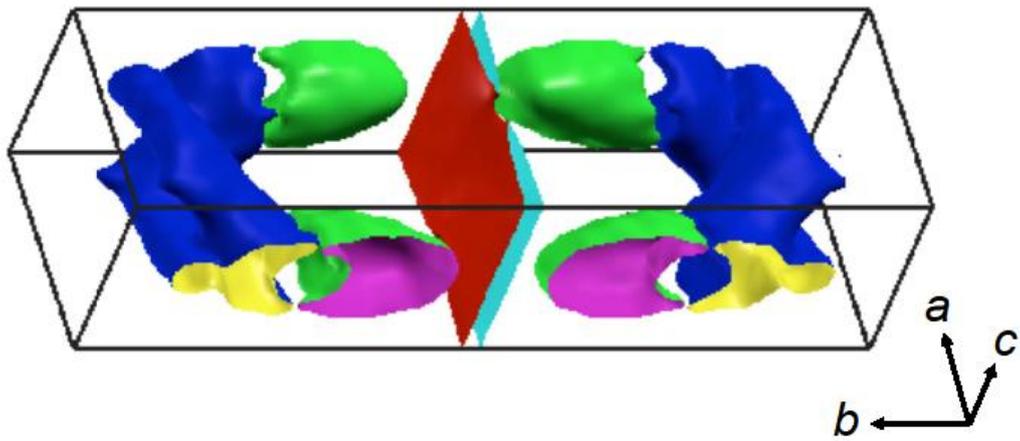

Figure 9. (Color online) Calculated Fermi surfaces for Nb$_2$PdSe$_5$ in a conventional reciprocal unit cell. The Fermi surfaces consist of the electron surfaces of sheet (red and cyan) and distorted closed pockets (green) and the hole surfaces of corrugated cylinder (blue).